\documentclass[%
preprint,
amsmath,amssymb,
]{revtex4-2}
\usepackage{amsmath}
\usepackage{graphicx}
\usepackage{dcolumn}
\usepackage{bm}
\usepackage{lineno} 
\DeclareMathOperator\erf{erf}

\begin{document}

\title{Photoacoustic effect from an oscillating source in a one-dimensional resonator}
\author{Xingchi Yan}
\email{xingchi\_yan@brown.edu}
\author{Siyuan Song}
\author{Gerald J. Diebold}
\affiliation{
 Department of Chemistry, Brown University, Providence, RI, 02912, USA
}
\affiliation{
 School of Engineering, Brown University, Providence, RI, 02912, USA
}


\begin{abstract}
Although the photoacoustic effect is most commonly generated by pulsed or amplitude modulated continuous optical sources, it is possible to generate acoustic waves by moving a constant amplitude, continuous light beam. If the light beam moves at the speed of sound, an amplification effect takes place which can be used in trace gas detection. Here, the properties of the photoacoustic effect are investigated for a continuous optical beam moving in a one-dimensional resonator. The solution shows the additive effects of sweeping the optical beam the length of the cell and back. 

\end{abstract}

\maketitle


\section{Introduction}

Photoacoustic trace gas detection has a rich history dating from the 1960's\cite{kreuzer1971ultralow, dewey1973acoustic, lyamshev2004radiation, lasersinacoustics}. 
The research on this subject\cite{hodgkinson2012optical, hess1983resonant, sigrist1994air,sigrist1992environmental,sigrist1998air, brand1995pulsed, zharov2013laser,claspy1977infrared, elia2009photoacoustic, karbach1985high, karbach1986photoacoustic}, has focused on optical excitation schemes and the design of the resonator with the goal of providing the highest possible sensitivity. The photoacoustic effect's amplitude exhibits unbounded linear growth over time when an optical source moves at the speed of sound in a one-dimensional geometry within the linear acoustic regime \cite{gk, xiong2017photoacoustic, bai2019moving}. Recently, the use of a moving infrared optical grating has been employed to generate the photoacoustic effect in the hundreds of kilohertz range in a cavity with a resonant piezoelectric crystal used as the detector \cite{xiong2017photoacoustic}. In this device a principle formulated by Gusev and Karabutov \cite{gk} where the optical source is made to move at the sound speed results in coherent addition of the instantaneously produced light induced pressure wave with the traveling pressure wave generated at a previous time was used to advantage. The reported detection limit for this device is in the parts per quadrillion range for $\text{SF}_6$, however, the instrument uses expensive technology and requires a crystal detector that is not at this point commercially available. As a result, it was decided to investigate the principles of operation of a longitudinal resonant cavity where a laser beam is directed to move back and forth along the axis of the resonator. Such an instrument would employ a microphone operating in the range of kHz instead of the much higher frequency reported in ref. \cite{xiong2017photoacoustic}. 

For optically thin bodies irradiated by short light pulses, solutions to the photoacoustic wave equations \cite{morse1986theoretical,westervelt1973laser, diebold1991photoacoustic, calasso2001photoacoustic} have been previously given in a list of references. These solutions, depending on the geometries such as spheres \cite{diebold1988photoacoustic,diebold1990photoacoustic,yan2021generation}, layers \cite{khan1993photoacoustic}, and cylinders \cite{khan1996photoacoustic}, show distinctive temporal features that permit applications in sensing and imaging  \cite{wang2017photoacoustic}.
Here we give solutions to the wave equation for the photoacoustic effect generated by oscillating sources in a one-dimensional cell. Sequentially, the paper gives the solutions to the wave equation for a single delta source, a moving delta source, and a moving Gaussian light source in a one-dimensional resonator in each respective section. The concluding section discusses the results and outlines directions for future research.

\section{Single delta pulse in a 1D resonator}\label{single_delta_sec}
The motion of a constant amplitude, continuous laser beam along a longitudinal resonator can be modeled as a summation of delta functions at various $z_n$ which are the coordinates of each moving delta source. 

Consider a single moving delta heat source in a one-dimensional resonator that extends from 0 to $L$ irradiated with a laser beam, the heating function can be written as the form $H=\delta(z-ct)=\frac{I}{c}\delta(t-z/c)$ where $I$ is the energy per unit area and time delivered to the gas and $z$ is the coordinate along the symmetry axis of the resonator. The photoacoustic effect in a inviscid fluid is described by \cite{gk, westervelt1973laser},
\begin{equation}
\left( \nabla ^{2}-\frac{1}{c^{2}}\frac{\partial ^{2}}{\partial t^{2}}
\right) p=-\frac{\beta }{C_{P}}\frac{\partial H}{\partial t}, \label{WE1}
\end{equation}
where $p$ is the pressure, $c$ is the sound speed, $\beta$ is the thermal expansion coefficient, $C_p$ is the specific heat capacity, $t$ is the time, and $H$ is the heating function, which describes the energy per unit time and volume deposited 
by the optical source. The one-dimensional photoacoustic wave equation can be written as,
\begin{equation}
   \Bigg(\frac{\partial ^2}{\partial z^2}-\frac{1}{c^2}\frac{\partial^2}{\partial t^2}\Bigg)p=-\frac{\beta I}{C_pc}\frac{\partial}{\partial t}\delta(t-z/c).
    \label{single_delta}
\end{equation}

\begin{figure}
\includegraphics[width=0.65\textwidth]{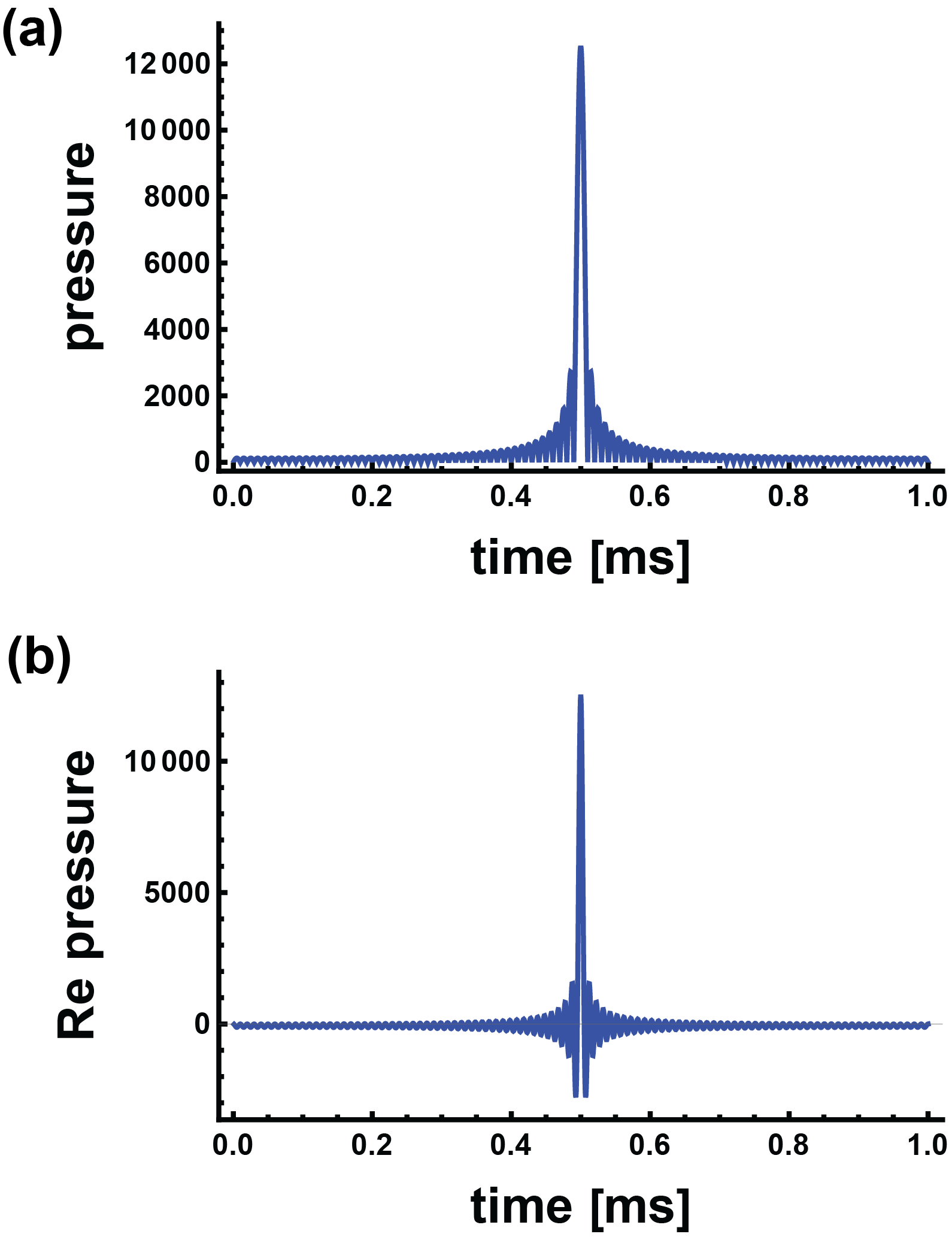}
\caption[Time domain pressure for a single delta pulse in a one-dimensional resonator in short time interval (0, 0.001 s)]{Time domain pressure for single delta pulse in a one-dimensional resonator in short time interval (0, 0.001 s) (a) pressure, (b) real part of pressure. The parameters used are $z=\frac{L}{2}$, $c=100$ m/s, $L=0.1$ m, $\beta=1$, $C_p=1$, $I=1$ and $M=200$.}
\label{single_delta_pic}
\end{figure}

\begin{figure}[!ht]
\includegraphics[width=0.65\textwidth]{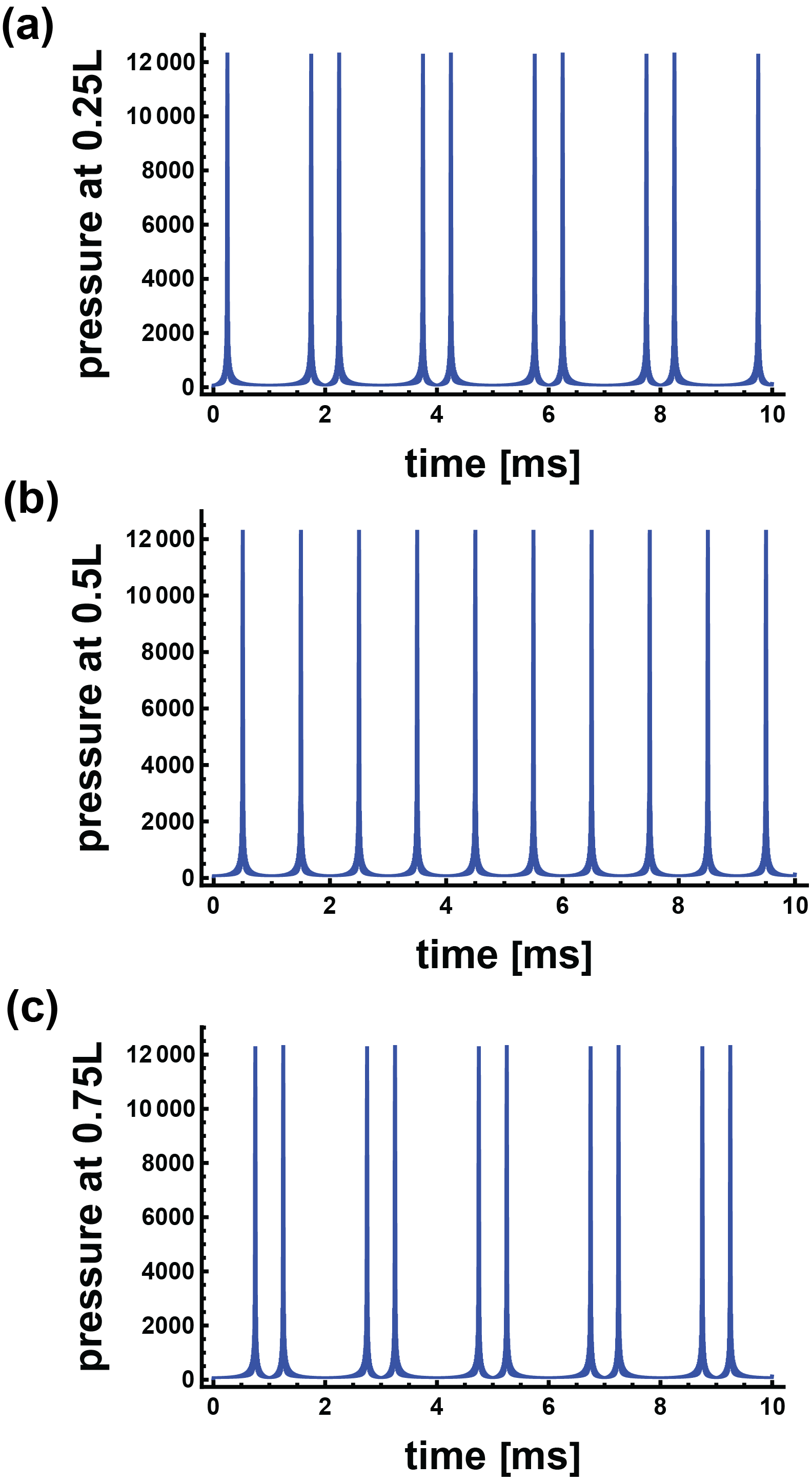}
\caption[Time domain pressure for single delta pulse in a one-dimensional resonator in longer time interval (0, 0.01 s)]{Evaluation of the time domain pressure for single delta pulse in a one-dimensional  resonator in a long time interval (0, 0.01 s). The parameters taken are (a) $z=\frac{L}{4}$, (b) $z=\frac{L}{2}$, (c) $z=\frac{3L}{4}$, $c=100$ m/s, $L=0.1$ m, $\beta=1$, $C_p=1$, $I=1$ and $M=200$.}
\label{singe_delta_pro}
\end{figure}

Fourier transform of eq. \ref{single_delta} followed by integration of the source term by parts gives a Helmholtz equation for the frequency domain pressure $\tilde{p}$ as, 
\begin{equation}
\begin{split}
   \Bigg(\frac{\partial ^2}{\partial z^2}+\frac{\omega^2}{c^2}\Bigg)\tilde{p}=-\frac{\beta I}{C_pc}\int_{-\infty}^{\infty}e^{-i\omega t}\frac{\partial}{\partial t}\delta(t-z/c)dt \\=-\frac{i \omega \beta I}{C_pc}\int_{-\infty}^{\infty}e^{-i\omega t}\delta(t-z/c)dt=-\frac{i \omega \beta I}{C_pc}e^{-i\omega z/c}.
    \label{single_equ}
\end{split}
\end{equation}
A Green's function \cite{mathews1970mathematical, arfken2005mathematical} for the one-dimensional resonator is given as, 
\begin{equation}
   G(z,z')=\frac{2}{L}\sum_{m=1}^{\infty}\frac{\cos k_mz \cos k_mz'}{k^2-k_m^2},
   \label{green_func_equ}
\end{equation}
where $k=\omega/c$ and $ k_m=\frac{m\pi}{L}$. Following an integration of eq. \ref{single_equ} over Green's function, the frequency domain pressure can be obtained as,
\begin{equation}
    \tilde{p}=-\frac{2i \omega \beta I}{C_pcL}\sum_{m=1}^{\infty}\frac{\cos(k_mz)}{k^2-k_m^2}\int_0^Le^{-i\omega z'/c}\cos(k_mz')dz'.
    \label{green_1d}
\end{equation}
The integral in eq. \ref{green_1d}, denoted $K_0$, can be evaluated for the condition where $\sin(k_mL)=0$ and where $r=\cos(k_mL)=(-1)^m$,
\begin{equation}
   K_0= \int_0^Le^{-i\omega z'/c}\cos(k_mz')dz'
   =\frac{-i\omega}{c(\frac{\omega^2}{c^2}-k_m^2)}\Big(1-e^{-i\omega L/c}r\Big).
   \label{K0_equ}
\end{equation} 
Inverse Fourier transformation of eq. \ref{green_1d} gives the time domain photoacoustic pressure as, 
\begin{equation}
   p(t) = \frac{-\beta Ic}{LC_p}\sqrt{\frac{\pi}{2}}\sum_{m=1}^M\frac{\cos(k_mz)}{k_m}e^{ick_mt}\Bigg[i-ck_mt+e^{iLk_m}(-1)^m(-i+k_mL+ck_mt)\Bigg].
   \label{single1}
\end{equation}
For values of the parameters $z=L/2$, $c=100$ m/s, $L=0.1$ m, $\beta=1$, $C_p=1$, $I=1$ and $M=200$, eq. \ref{single1} gives the pressures
plotted in fig. \ref{single_delta_pic} and fig. \ref{singe_delta_pro}. The real and imaginary parts of the solution for a single delta function pulse source in the time interval of $(0,0.001 \text{s})$ are shown in fig. \ref{single_delta_pic}. For longer time intervals $(0,0.01 
\text{s})$ the response is shown in fig.  \ref{singe_delta_pro}.  Note that for $z=L/2$, the pressure peaks are observed at 0.5 ms, 1.5 ms, 2.5 ms, 3.5 ms, etc. as shown in fig. \ref{singe_delta_pro} (b). When $z=L$, the pressure peaks are observed at 1 ms, 3 ms, 5 ms, 7 ms, etc. A similar mathematical treatment can be applied for the other delta sources, for example, for a delta source with $z_1=L-c(t-L/c)=2L-ct$. Results for a summation of single delta sources
are given immediately below.

\section{Moving optical sources in a 1D resonator \label{multi_delta_section} }

Consider a one-dimensional resonator that extends from 0 to $L$ irradiated with a laser beam that repeatedly moves the length of the resonator and then reverses direction. For the present problem the heating function is taken to be
\begin{equation}
     H(z)=I\sum_n\delta(z-z_n), \label{delta_heating}
\end{equation}
where $I$ is the energy per unit area and time delivered to the gas, $z$ is the coordinate along the symmetry axis of the resonator, and $z_n$ is the coordinate of each of the moving delta function sources. The motion of the laser beam along the $z$ axis is described by the arguments of the delta functions containing the various $z_n$. Table \ref{argument_delta1} gives values of $z-z_n$ and the corresponding delta function (absent the factor $c$ in the denominators in the right hand column) to be used in eq. \ref{delta_heating} for the first few values of $n$, where the identity $\delta(a-ct)=\delta(t-a/c)/c$ has been used. 

\begin{table}
\begin{center}
\begin{tabular}{ |c|c|} 
 \hline
 $z-z_0=z-ct$ & $\delta(t-\frac{z}{c})$  \\ 
 \hline
 $z-z_1=z-2L+ct$  & $\delta(t-[\frac{2L}{c}-\frac{z}{c}])$  \\ 
 \hline
 $z-z_2=z+2L-ct$ &  $\delta(t-[\frac{2L}{c}+\frac{z}{c}])$   \\ 
 \hline
 $z-z_3=z-4L+ct$ &  $\delta(t-[\frac{4L}{c}-\frac{z}{c}])$   \\ 
 \hline
 $z-z_4=z+4L-ct$ & $\delta(t-[\frac{4L}{c}+\frac{z}{c}])$    \\ 
  \hline
\end{tabular}
\caption{ \label{argument_delta1} Arguments of the delta functions in Eq. \ref{delta_heating}} 
\end{center}
\end{table}

The first delta function originates at the left end of the resonator and moves to the right along the $z$ axis. At the time $t = L / c$, the $z_1$ delta function begins moving from the right end of the resonator traveling to the origin. Thus all of the even numbered entries move from left to right while the odd numbered ones move from the right end of the resonator to the left. 

Fourier transformation of eq. \ref{WE1} followed by integration of the source term by parts gives a Helmholtz equation for the frequency domain pressure $\tilde{p}$ as
\begin{equation}
  \Big(\frac{\partial^2}{\partial z^2}+\frac{\omega^2}{c^2}\Big)\tilde{p}=-\frac{i\omega \beta I}{C_pc}\int_{-\infty}^{\infty}\sum_{n=0}\delta(t-t_n)e^{-i\omega t}dt,
  \label{helmholtz_delta}
\end{equation}
where $t_n=(nL+z)/c$ for even $n$ and $t_n=((n+1)L-z)/c$ for odd $n$.

Following an integration of eq. \ref{helmholtz_delta} over the Green's function, the frequency domain pressure can be found as follows: 
for even $n$ the wave equation becomes,
\begin{equation}
    \Big(\frac{\partial^2}{\partial z^2}+\frac{\omega^2}{c^2}\Big)\tilde{p}=-\frac{i\omega \beta I}{C_pc}e^{-i\frac{n\omega L}{c}}K_{\text{even}},
    \label{Ke}
\end{equation}
whereas for odd it is,
\begin{equation}
    \Big(\frac{\partial^2}{\partial z^2}+\frac{\omega^2}{c^2}\Big)\tilde{p}=-\frac{i\omega \beta I}{C_pc}e^{-i\frac{(n+1)\omega L}{c}}K_{\text{odd}},
     \label{Ko}
\end{equation}
where $K_{\text{even}}=\displaystyle \int_{0}^{L} \cos (k_mz')e^{-i\omega z'/c}dz'$ as evaluated in eq. \ref{K0_equ}, $K_{\text{odd}}=K_{\text{even}}^*$ and $r=(-1)^m$. The frequency domain photoacoustic pressure for even $n$ becomes,
 \begin{equation}
       \tilde{p}=-\frac{2\omega^2\beta I}{C_pc^2L}\sum_{n,m}\frac{\cos(k_mz)}{\Big(\omega^2/c^2-k_m^2\Big)^2}e^{-in\omega L/c}\Bigg(1-re^{-i\omega L/c}\Bigg),
       \label{even}
\end{equation}
while for odd $n$ it is
 \begin{equation}
 \tilde{p}=\frac{2\omega^2\beta I}{C_pc^2L}\sum_{n,m}\frac{\cos(k_mz)}{\Big(\omega^2/c^2-k_m^2\Big)^2}e^{-i(n+1)\omega L/c}\Bigg(1-re^{i\omega L/c}\Bigg);
       \label{odd}
\end{equation}
combining the two expression, it becomes, 
\begin{equation}
    \tilde{p}=-\frac{2\omega^2 \beta I}{C_pc^2L}\sum_{n,m}\frac{x\cos(k_mz)}{(\omega^2/c^2-k_m^2)^2}e^{-i\omega(\frac{Ln}{c}+\frac{L(x-1)}{2xc})}\Big(1-re^{-ix\omega L/c}\Big).
\end{equation}
where $x=(-1)^n$.

Inverse Fourier transformation  of eq. \ref{even} and eq. \ref{odd} can be obtained using Mathematica, which gives, 
\begin{equation}
    p=\frac{-\beta Ic}{LC_p}\sqrt{\frac{\pi}{2}}\sum_{n,m}\frac{\cos( k_mz)}{k_m}\Theta(t-nL/c)(-1)^{n+1}\bar{p},
   \label{delta}
\end{equation}
where $\Theta$ is the Heaviside function and $\bar{p}$ in eq. \ref{delta} is defined as,
\begin{multline}
    \bar{p}=ck_mr(a-b+t)\cos\Big(ck_m(a-b+t)\Big)-\sin\Big(ck_m(a+t)\Big)\\ -ck_m(a+t)\cos \Big(ck_m(a+t)\Big)+r\sin\Big(ck_m(a-b+t)\Big),
\end{multline}
where $a$ and $b$ are defined as for even $n$: $a=nL/c$ and $b=-L/c$; and for odd $n$: $a=(n+1)L/c$ and $b=L/c$. The summation in eq. \ref{delta} over $n$ refers the number of sweeps of the optical beam, beginning at 0; the summation over $m$ starts at 1, and in principle extends to infinity. 

\begin{figure}[!ht]
\includegraphics[width=0.65\textwidth]{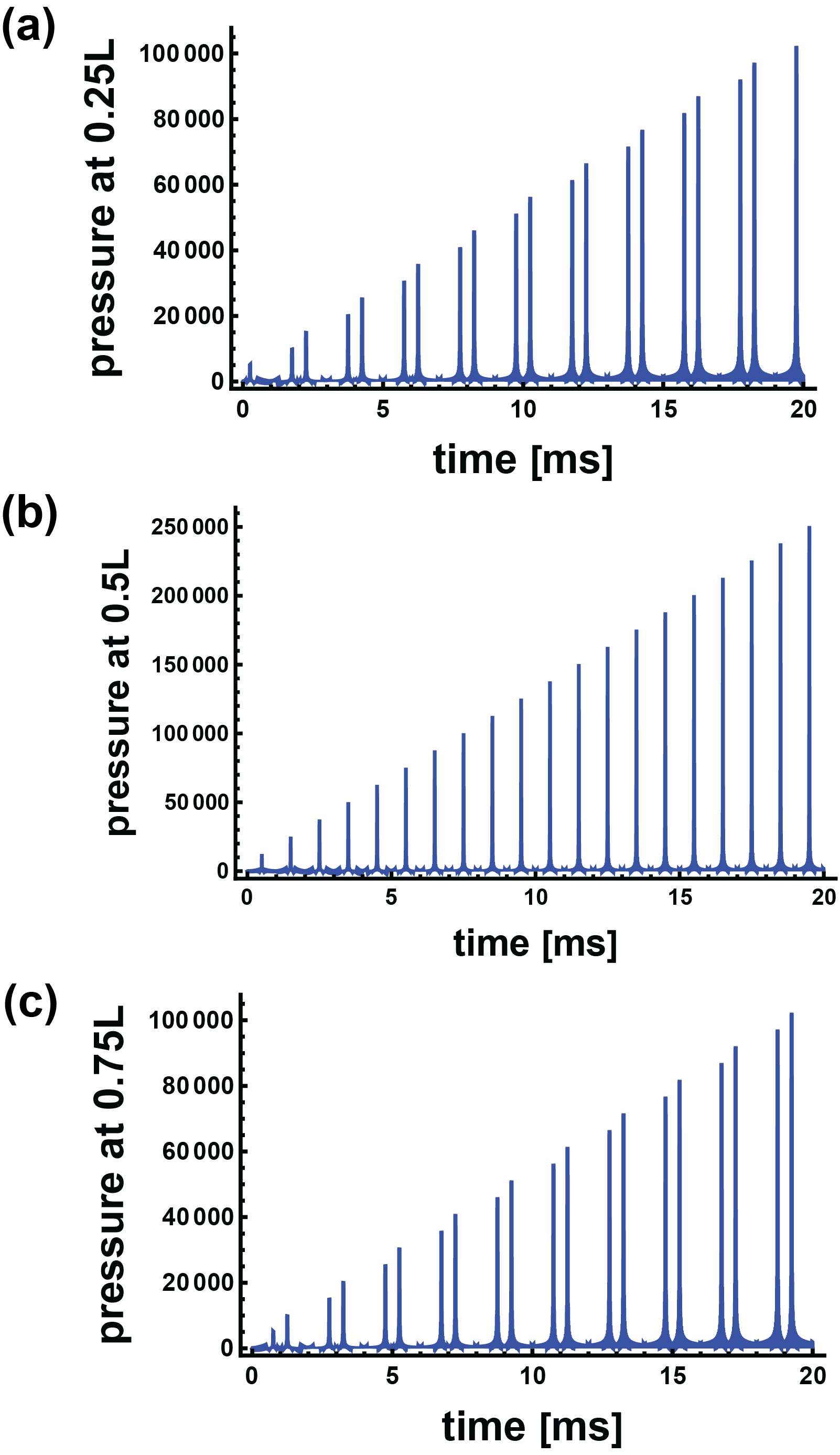}
\caption{Time domain pressure for an oscillating delta pulse in a one-dimensional resonator. Magnitude of the photoacoustic pressure $p$ versus time in milliseconds from eq. \ref{delta} evaluated at (a) $z=\frac{L}{4}$, (b) $z=\frac{L}{2}$ and $z=\frac{3L}{4}$ of the resonator for $N=50$, $M=200$, $L=0.1$ m and $c=100$ m/s. }
\label{osi_delta_fig}
\end{figure}

For values of the parameters  $c=100$ m/s, $L=0.1$ m, $\beta=1$, $C_p=1$, $I=1$, $N=50$, $M=200$, eq. \ref{delta} gives the pressures as
plotted in fig. \ref{osi_delta_fig}.  Note that for $z=L/2$, the pressure peaks are observed at 0.5 ms, 1.5 ms, 2.5 ms, 3.5 ms, etc. as shown in fig. \ref{osi_delta_fig} (b). The amplitude of the pressure increases linearly with the number of
sweeps across the resonator.

\section{Oscillating Gaussian \label{moving_gaussian} }

Consider an oscillating Gaussian source with speed $v$ along the longitudinal axis. The heating function can be written as $H=\frac{I_0}{\sigma \sqrt{\pi}}\sum_ne^{-(\frac{z-z_n}{\sigma})^2}$. Fourier transformation of eq. \ref{WE1} followed by integration of the source term by parts gives the frequency domain pressure $\tilde{p}$ as,
\begin{equation}
\Big(\frac{\partial^2}{\partial z^2}+\frac{\omega^2}{c^2}\Big)\tilde{p}=-\frac {i\omega \beta I}{C_p\sigma \sqrt{\pi}}\int_{-\infty}^{\infty}e^{-i\omega t}\sum_{n}e^{-(\frac{z-z_n}{\sigma})^2}dt,
\label{pa_osci_eq}
\end{equation}
where the coordinate of each of the moving Gaussian source is $z_n=\frac{1}{2}L(1-x)+xv(t-\frac{nL}{v})$. Through use of $z-z_n=-xv(t-t_0)$ and $t_0:=\frac{z}{xv}+\frac{L}{2xv}(x-1)+\frac{nL}{v}$ (See Appendix), the integral in eq. \ref{pa_osci_eq} becomes,
\begin{equation}
    \int_{-\infty}^{\infty}e^{-i\omega t}\exp\Big(-\frac{x^2v^2}{\sigma^2}(t-t_0)^2\Big)dt=\frac{\sqrt{\pi}\sigma}{v}\exp\Big(-\frac{\omega^2\sigma^2}{4v^2}-i\omega t_0\Big).
\end{equation}
Evaluation of eq. \ref{pa_osci_eq} at $v=c$ using the Green’s function in eq. \ref{green_func_equ} gives the frequency domain pressure as, 
\begin{equation}
    \tilde{p}=-\frac{2i\omega \beta I}{C_pLc}\sum_{n,m}\frac{\cos(k_mz)}{k^2-k_m^2}e^{-\frac{\omega^2\sigma^2}{4c^2}-i\omega(\frac{Ln}{c}+\frac{L(x-1)}{2xc})}\int_{0}^{L}\cos(k_mz')e^{-i\omega \frac{z'}{xc}}dz'.
     \label{pnew1} 
\end{equation}
Evaluation of the integral in eq. \ref{pnew1} gives, 
\begin{equation}
\int_{0}^{L}\cos(k_mz')e^{-i\omega \frac{z'}{xc}}dz'=\frac{-i\omega x}{c}\frac{[1-r\exp(-i\omega L/(xc))]}{(\omega/c)^2-k_m^2},
\end{equation}
where $x=(-1)^n$ and $r=(-1)^m$.  
Hence, the photoacoustic pressure becomes,
\begin{equation}
    \tilde{p}=-\frac{2\beta I}{C_pc^2L}\sum_{n,m}\frac{x\cos(k_mz)\omega^2}{(\omega^2/c^2-k_m^2)^2}e^{-\frac{\omega^2\sigma^2}{4c^2}}e^{-i\omega(\frac{Ln}{c}+\frac{L(x-1)}{2xc})} \Big(1-re^{-ix\omega L/c}\Big).
\end{equation}
Inverse Fourier transform of $F(\omega)=\frac{\omega^2}{(\omega^2/c^2-k_m^2)^2}e^{-\frac{\omega^2\sigma^2}{4c^2}}$ (see Appendix) gives, 
\begin{equation}
\begin{split}
    f(t)=\frac{c^2}{2k_m^2}\Bigg(e^{-ick_mt}g^{[2]}(t;ck_m)+e^{ick_mt}g^{[2]}(t;-ck_m) \\
    -\frac{1}{ck_m}\Big(e^{-ick_mt}g^{[1]}(t;ck_m)-e^{ick_mt}g^{[1]}(t;-ck_m)\Big)\Bigg),
    \label{factor_FT}
\end{split}
\end{equation}
where the functions $g^{[1]}(t;ck_m)$ and $g^{[2]}(t;ck_m)$ are given in the 
Appendix. 

The other terms contributes as a phase shift thus the time domain photoacoustic pressure becomes,
\begin{equation}
\begin{aligned}
    p(t) &= \frac{-2\beta I}{C_pc^2L}\sum_{n,m}x\cos(k_mz) \Theta(t-nL/c) \\
   &\hspace{-1.5em}\times \Bigg(f\Big(t+\frac{Ln}{c}+\frac{L(x-1)}{2xc}\Big)-rf\Big(t+\frac{Ln}{c}+\frac{L(x-1)}{2xc}+\frac{xL}{c}\Big)\Bigg)
    \label{gaussian_equ}
\end{aligned}
\end{equation}

In the limit $\sigma \rightarrow 0$, the solution for oscillating Gaussian eq. \ref{gaussian_equ} reduces to the solution for oscillating delta eq. \ref{delta}.

\begin{figure}[!ht]
\includegraphics[width=0.65\textwidth]{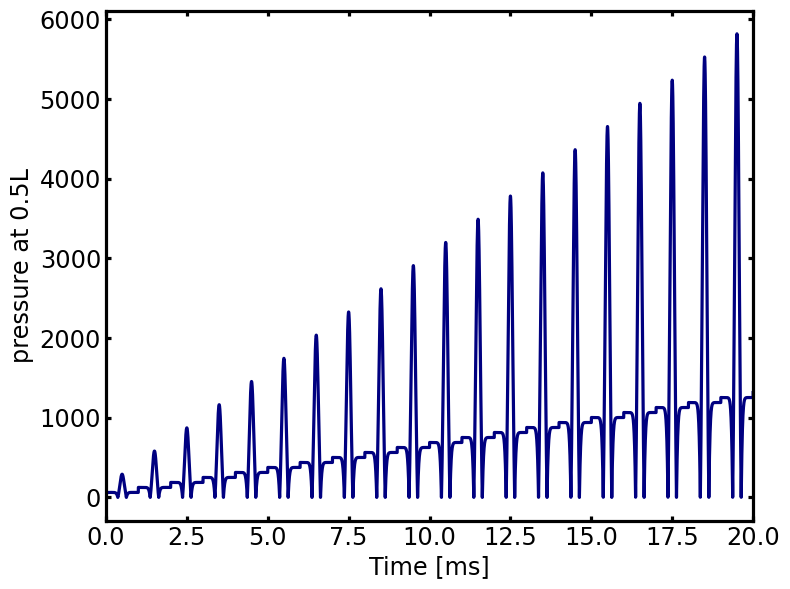}
\caption{Time domain pressure for an oscillating Gaussian source in a
one-dimensional resonator. Magnitude of the photoacoustic
pressure $p$ versus time in milliseconds from eq. \ref{gaussian_equ} at $z=\frac{L}{2}$ of the resonator for $N=50$, $M=150$, $L=0.1$ m, and $c=100$ m/s. }
\label{gaussian_plot}
\end{figure}

For values of the parameters  $c=100$ m/s, $L=0.1$ m, $\beta=1$, $C_p=1$, $I=1$, $N=50$, $M=150$, $\sigma=0.01$, and $z=L/2$, eq. \ref{gaussian_equ} gives the pressures
plotted in fig. \ref{gaussian_plot}.  The pressure peaks are observed at 0.5 ms, 1.5 ms, 2.5 ms, 3.5 ms, etc. as shown in Fig. \ref{gaussian_plot}.

\section{Discussion}
\label{diss}

The results of the calculations presented here give an expression for the
photoacoustic pressure generated by a moving light beam in an optically
thin absorbing fluid in a one-dimensional resonator. It can be seen that the amplitude of the pressure increases linearly with the number of sweeps across the resonator. This addition is reminiscent of pulsed excitation of a conventional resonator when the time between pulses corresponds to a resonance of the photoacoustic cell. Such pulse addition has been used to advantage in ref. \cite{brand1995pulsed}
where a cylindrical resonator containing infrared absorbing gas was excited by a pulsed laser. 
Both the pulsed excitation and moving source have in common
synchronization of excitation, however, only the latter gives the
amplification that appears to be unique in that the excitation takes place additively at all times.

Of note is that only the case of a source that instantaneously switches
direction as it approaches the ends of the cavity has been considered,
which, in practice may present a difficulty requiring a large bandwidth for
the optical modulator that provides the directional modulation. A simpler
experimental arrangement might employ sinusoidal motion in space, where
the speed of the optical source varies and does not match the sound speed
at all times. Such a configuration was not treated here, but it is clear that
rapid switching of the direction is optimal, which simplifies the mathematics
and shows the major features of a moving source photoacoustic effect in a
one-dimensional resonator. 

As the above calculations show, the amplification effect for a light source comoving with the sound speed as described in ref. \cite{gk} is extant in a cavity
for a source whose position instantaneously changes direction at the end
points of the resonator. It is clear from inspection of eq. \ref{pa_osci_eq} and eq. \ref{gaussian_equ} that the
pressure in the cavity is a summation of wavelets, with the final pressure coming as a result of acoustic interference. The results given here suggest a straightforward implementation of the amplification effect is possible for trace gas detection without the complicated moving grating design and
sophisticated detector as reported in ref. \cite{xiong2017photoacoustic}.  For other possible further investigation, it might be interesting to detect and engineer the spatial-temporal characteristics of the emitted acoustic waves \cite{gk, diebold1990photoacoustic, diebold1991photoacoustic, yan2021abel} through multiscale modeling \cite{weinan2011principles} , machine learning \cite{ jin2023recent, pmlr-v202-molinaro23a} and experiments, given extensive 
literature in model-based and data-driven approaches for inverse design of mechanical systems \cite{jin2022dynamic, jin2021ruga}. 

Regarding experimental implementation of the moving source concept developed here, construction of a resonator with an audio frequency microphone is straightforward.  A moving source can be implemented using an electro-optic oscillating mirror, which would yield sinusoidal or some variant of sinusoidal motion in the resonator depending on the design of the modulator, or, alternately, through use of a wide bandwidth acoustic-optic modulator which could provide near perfect matching of the motion of the optical beam to the motion of the acoustic wave at all times. 

\section{Appendix}

\renewcommand{\theequation}{A\arabic{equation}} 
\setcounter{equation}{0} 

The following identity has been used in the evaluation of eq. \ref{pa_osci_eq}, 
\begin{equation}
z-z_n=z-\frac{1}{2}L(1-x)-xvt+\frac{xvnL}{v}
=-xv\Big[t-\Big(\frac{z}{xv}+\frac{L}{2xv}(x-1)+\frac{nL}{v}\Big)\Big]=-xv(t-t_0),
\end{equation}
where $t_0=\frac{z}{xv}+\frac{L}{2xv}(x-1)+\frac{nL}{v}$. 

\subsection{Derivation Eq. \ref{factor_FT}: the forms of $g^{[1]}$ and $g^{[2]}$}

Rewrite $F(\omega)$ as,
\begin{equation}
\begin{aligned}
F(\omega) &= \frac{\omega^2}{(\omega^2/c^2-k_m^2)^2}e^{-\frac{\omega^2\sigma^2}{4c^2}} \\
&= \frac{c^2\omega^2}{2k_m^2}e^{-\frac{\omega^2\sigma^2}{4c^2}}\Bigg(\frac{1}{(\omega-ck_m)^2}+\frac{1}{(\omega+ck_m)^2} -\frac{1}{ck_m(\omega-ck_m)}+\frac{1}{ck_m(\omega+ck_m)}\Bigg).
\end{aligned}
\end{equation}

The Fourier transform of each term can be evaluated separately, which gives Eq. \ref{factor_FT} where 
\begin{equation}
    g^{[1]}(t;ck_m)= \frac{2\sqrt{2}c^3}{\sigma^3}\Big(-it+\frac{k_m\sigma^2}{2c}\Big)e^{ick_mt-t^2c^2/\sigma^2}+
    \frac{c^2k_m^2i\sqrt{\pi}}{\sqrt{2}}e^{-\frac{k_m^2\sigma^2}{4}}\erf\Bigg({\frac{-ct+ik_m\sigma^2/2}{\sigma}}\Bigg)
\end{equation}
\begin{equation}
\begin{aligned}
    g^{[2]}(t;ck_m) &= \frac{\sqrt{2}}{\sigma/c}e^{ick_mt-t^2c^2/\sigma^2} \\
    &\quad + ck_me^{-k_m^2\sigma^2/4}\sqrt{\frac{\pi}{2}} \Bigg(-ck_mt+\frac{ik_m^2\sigma^2}{2}+2i\erf\Bigg(\frac{-ct+ik_m\sigma^2/2}{\sigma}\Bigg)\Bigg)
\end{aligned}
\end{equation}
where erf is the error function. 

\subsection{Numerical evaluation of the theoretical solution}

Eq. \ref{gaussian_equ} is the general solution to the Gaussian case. For any fixed $n$ and $m$, the Fourier transform of Eq. \ref{pnew1} can be compared to the theoretical solution in Eq. \ref{gaussian_equ}. The singularity term in the denominator will lead to instability in evaluating of fast Fourier transform (FFT). The singular term and the non-singular term can be sepatated into $\tilde{p}=\tilde{p}_{\text{singular}}+\tilde{p}_{\text{non-singular}}$. For the singular term, there is an analytical Fourier transform formula. 
Evaluation of the Fourier transform of the singular term gives,
\begin{equation}
    F\Bigg[\frac{\omega^2}{(\omega^2/c^2-k_m^2)^2}\Bigg]=\frac{-c^3\sqrt{\pi}}{2\sqrt{2}k_m}\Bigg(ck_mt\cos(ck_mt)+\sin(ck_mt)\Bigg)
    \label{singular_term}
\end{equation}

\begin{figure}[ht!]
\includegraphics[width=0.6\textwidth]{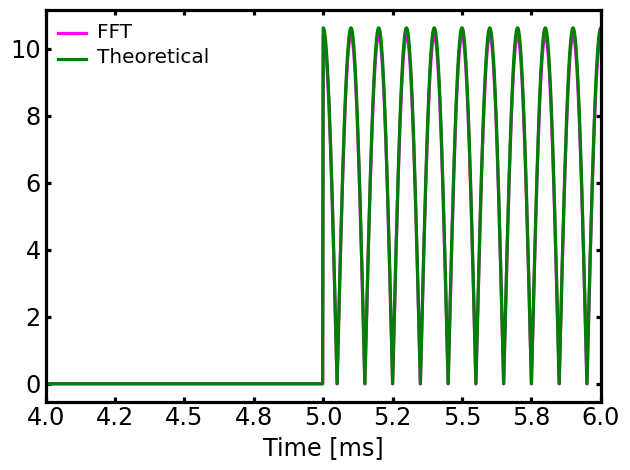}
\caption{Numerical Evaluation of the term in eq. \ref{gaussian_equ} with $n=5$ and $m=10$ via both the theoretical expression and FFT evaluation of the term in eq. \ref{pnew1}.}
\label{SI_check}
\end{figure}

Evaluation of the theoretical expression in Eq. \ref{gaussian_equ} with $n=5$ and $m=10$ and numerical FFT of the non-singular term, adding the theoretical evaluation of the singular term eq. \ref{singular_term}, gives the pressure as shown in Fig. \ref{SI_check}. The FFT can be seen to align precisely  with the theoretical solution described in Eq. \ref{gaussian_equ}. 

\newpage
\bibliography{apssamp}

\end{document}